\documentstyle[aps,prl,twocolumn,psfig]{revtex}

\begin{document} 


\twocolumn[\hsize\textwidth\columnwidth\hsize\csname@twocolumnfalse\endcsname 
\title{\bf Andreev Reflection Enhanced Shot Noise in Mesoscopic SNS Junctions} 
\author{X. Jehl, P. Payet-Burin, C. Baraduc, R. Calemczuk and M. Sanquer} 
\address{ CEA-DSM-DRFMC-SPSMS,  CEA-Grenoble \\  38054 Grenoble Cedex 9, France.} 
\maketitle 	
\begin{abstract} 
{Current noise is measured with a SQUID in low impedance and transparent Nb-Al-Nb junctions of length comparable to the phase breaking length and  much longer than the thermal length. The shot noise amplitude is compared with theoretical predictions of doubled shot noise in diffusive  normal/superconductor (NS) junctions due to Andreev reflections. We discuss the heat dissipation away from the normal part through the NS interfaces. A weak applied  magnetic field reduces the amplitude of the $1/f$ noise by a factor of two, showing that even  far from equilibrium the sample is in the  mesoscopic regime.}
\end{abstract}] 


\par Nonequilibrium noise in SNS junctions has been recently addressed experimentally \cite{dieleman}\cite{strunk}\cite{kozhevnikov}. Interest in this field has been motivated by the celebrated shot noise results obtained in short conductors connected to normal reservoirs \cite{steinbach}, in a two-dimensional electron gas \cite{liefrink} or in fractional quantum hall liquids \cite{glattli}\cite{reznikov}. The analysis of the shot noise amplitude as well as the crossover from the Johnson-Nyquist to the  shot noise regime provides information about the nature of the carriers beyond what is deduced from linear conductance measurements. It has been predicted (but not shown experimentally) that the shot noise in a mesoscopic normal diffusive sample connected to a superconducting reservoir at one end is doubled compared to the case of two normal reservoirs \cite{khlus}\cite{beenakker}. This reflects that at low temperature and low energy the charge transport is dominated by Andreev processes transferring electrons by pairs. Beyond the SN case, the nature of charge carriers in the SNS case is also a major issue both theoretically and experimentally. In the case of multiple Andreev reflections (MAR) theoretical works predict an excess current noise \cite{bezuglyi}\cite{naveh2}\cite{cuevas}.
\par Short SNS junctions have been studied by Dielemann {\it et al.} \cite{dieleman} in the case of pinholes in a NbN/MgO/NbN (SIS) structure. Below the superconducting gap, the shot noise they measure is much larger than expected for independent electrons. That is attributed to the coherent charge transfer of large multiple charge quanta. Hoss {\it et al.} \cite{strunk} have studied longer SNS junctions and found different types of behaviour depending on the value of the superconducting gap of electrodes: for large gap Nb electrodes, the quasiparticles are overheated, whereas for low gap Al electrodes a very large shot noise at low bias is attributed to the same mechanism as in ref. \cite{dieleman}.

\par
\begin{figure}
\psfig{figure=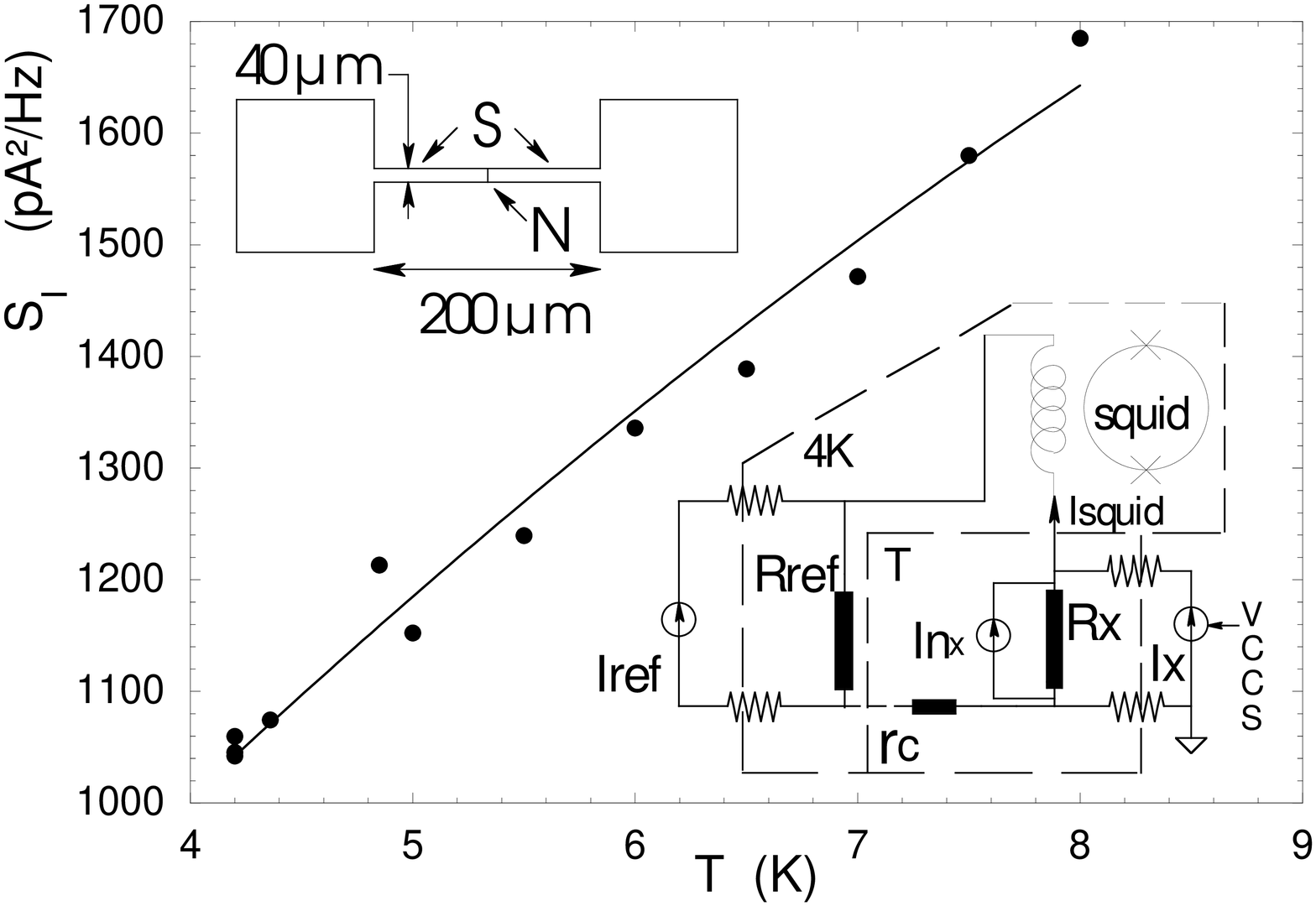,width=87mm}
\caption{Equilibrium noise in the $L=0.5\mu m$ sample. Solid curve gives the Johnson prediction with the measured temperature dependance of the sample resistance $R_{x}$. Right inset shows the resistance bridge and the SQUID. Only the current noise generator of interest ($i_{n_{x}}$) is represented. Typical values are $I_{ref}<10mA$, $R_{ref}=177m\Omega$, $r_{c}\approx 4m\Omega$, and $R_{x}\approx 0.25\Omega$. Left inset is a side view of the SNS structure.}
\label{fig1jehl}
\end{figure}

\begin{figure}
\psfig{figure=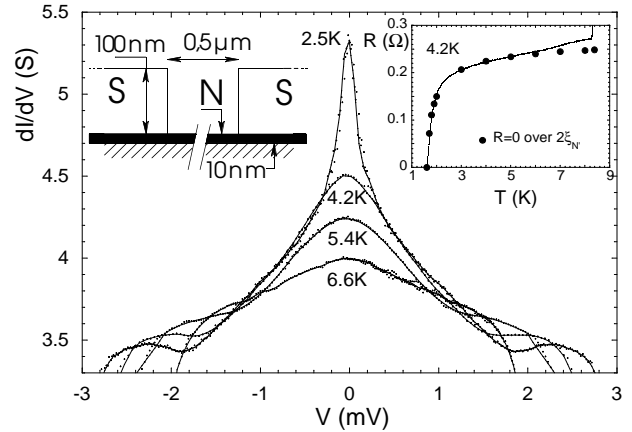,width=87mm}
\caption{Differential conductance dI/dV. The conductance peak reflects the proximity effect. Right inset shows the resistance versus temperature which is very much in agreement with calculations assuming a null resistance over twice the thermal length in aluminium (dots). Left inset shows a top view of the sample. }
\label{fig2jehl}
\end{figure}

 A SN junction with a low resistive noiseless normal reservoir at one side and a transparent SN interface at the other one requires several technological steps (e.g. multideposition and realignment). We fabricate a much simpler SNS junction which captures the same physics if the length of the junction is larger than the inelastic mean free path.  
We present shot noise measurements in Nb-Al-Nb junctions (above the critical temperature of aluminium) where the current noise is measured by a calibrated SQUID-based setup (Fig.~\ref{fig1jehl}) \cite{jehl}. 
In our high temperature range the sample length $L$ is much larger than the superconducting coherence length but comparable to the phase breaking length which is dominated by the electron-electron relaxation length $L_{ee}$.
Under these conditions the sample is in the mesoscopic regime where shot noise is only due to normal parts coherently attached to at least one of the superconducting reservoirs, but where MAR is inhibited ($L \approx L_{ee}$).
Indeed the conductance evolves in temperature and voltage as predicted for the standard proximity effect \cite{degennes}. The absence of conductance anomalies at finite bias (Fig.~\ref{fig2jehl}) indicates that Multi Particle Tunneling (i.e. coherent MAR process) is negligible. 
Our shot noise measurements show that the transport is indeed dominated by carriers whose effective charge is about twice that of the bare electron. At high temperature the shot noise is very much in agreement with the prediction for a diffusive normal metal connected to normal reservoirs ($S_{I}={2\over 3}e\langle I\rangle$)\cite{buttiker}, likely because the transport is mainly due to quasiparticles. But as the temperature decreases, the shot noise increases above this value. The evolution of the current noise power vs. bias current (including the crossover to the Johnson-Nyquist equilibrium noise) is consistent with an effective charge $2e$ at voltages well below the gap. In order to establish the role of carriers overheating in the noise properties in our SNS geometry \cite{strunk}, we have calculated the gradient of temperature produced at each SN interface by the Andreev thermal resistance and compared the resulting noise to the experimental data.

\par Another contribution to the noise is the $1/f$ noise. $1/f$ noise is found to be quantitatively in agreement with previous data. Its amplitude is reduced by a factor of two when a weak magnetic field is applied, as expected within the Feng-Lee-Stone theory of low-frequency resistance noise in dirty metals \cite{feng}\cite{birge}. Analysis of the field dependence shows that  $L_{\phi}$ is not substantially decreased even far from equilibrium.
\par Our SNS geometry as well as our temperature range differ from previous work. We start with a trilayer $10nm$ Al-$100nm$ Nb-$10nm$ Al made by sputtering in a single sequence on an $Si-SiO_{2}$ substrate. Then we define a mesa structure (upper inset in Fig. \ref{fig1jehl}) by optical lithography with a $200 \mu m \times 40 \mu m$ wire between large contact pads. The contact pads are further covered by a low resistance Ti-Au contact layer. By electron lithography and subsequent reactive ion etching we selectively etch the Nb-Al top layer over a length of $0.5 \mu m$ across the mesa wire (left inset in Fig. \ref{fig2jehl}). The resulting structure is a continuous $10nm$ thick Al layer, covered by two semi-infinite $100nm$-Nb layers separated by a gap of $0.5 \mu m \times 40 \mu m$. At $4.2K$ the 80 squares in parallel result in a resistance of $0.25\Omega$. The geometry is the inverse of the wire used in Ref. \cite{strunk}. The experiment is performed above the critical temperature of the aluminium film($1.6K$). We chose aluminium for the normal metal because of the good quality of the Al-Nb interface.

\par The current noise measurement scheme is based on a resistance bridge and a
 dcSQUID \cite{jehl} as shown in the inset of Fig. \ref{fig1jehl}. It is well adapted to our low impedance sample which has relatively high current noise but needs high bias currents to go beyond the thermal (equilibrium) noise regime. The bridge is composed of a reference resistance ($R_{ref}$) made with a macroscopic constantan wire, the sample ($R_{x}$) and the extra resistances in the superconducting loop ($r_{c}$) (dominated by the gold wires used to connect the sample). The current noise of the setup is $5pA/\sqrt{Hz}$. The total resistance of the bridge being $\approx 0.4 \Omega$, its Nyquist noise is $5.8 \times 10^{-22}A^{2}/Hz$ at $4.2K$ and is therefore more than 15 times bigger than the total noise of the electronic setup. A fit of the form $\alpha + \beta /f$ is always found in total agreement with the spectra for each value of the bias current \cite{jehl}, indicating two separable features: a $1/f$ component of amplitude $\beta$ and a white (i.e. frequency independent) noise level $\alpha$. Figure \ref{fig1jehl} shows the temperature dependance of the equilibrium noise. The solid line is the Johnson noise ($4kT/R_{x}$) calculated from the measured sample resistance. The data is very much in agreement with the prediction: the Nyquist noise is always recovered, thus showing the absolute calibration of the setup.

\par

Around $4.2K$ the temperature is much larger than the Thouless energy: the thermal length $L_{T} = \sqrt{\hbar D / k_{B}T}\simeq 0.08 \mu m $ (T=4.2K) is much shorter than both the sample length $L\simeq 0.5 \mu m$ and the phase breaking length $L_{\phi} \simeq 0.8 \mu m$ (T=4.2K). Both Josephson coupling and coherent MAR are negligible but $L$ is comparable to the electron-electron scattering length  $L_{ee} \simeq 1 \mu m$, and smaller than  the electron-phonon scattering length $L_{eph} \simeq 2.5 \mu m$, both estimated at $4.2K$. Therefore the shot noise is likely to be due to normal parts coherently attached to at least one superconducting reservoir.
The temperature dependence of the resistance exhibits two jumps corresponding to the two critical temperatures for niobium ($8.35K$) and aluminium ($1.6K$). Using the latter as the only parameter we can calculate the expected resistance by solving the equation for the coherence length in aluminium above its $T_{c}$ which differs from the thermal length \cite{degennes}. The result fits the data remarkably well (see right inset in Fig. \ref{fig1jehl}). The differential conductance (Fig. \ref{fig2jehl}) exhibits a peak which is another signature of this proximity effect. We also performed magnetoconductance measurements from which we inferred $L_{\phi} \approx 0.8 \mu m$ at $4.2K$, in quantitative agreement with previous data on aluminium films \cite{gershenson}.

\par
\begin{figure}
\psfig{figure=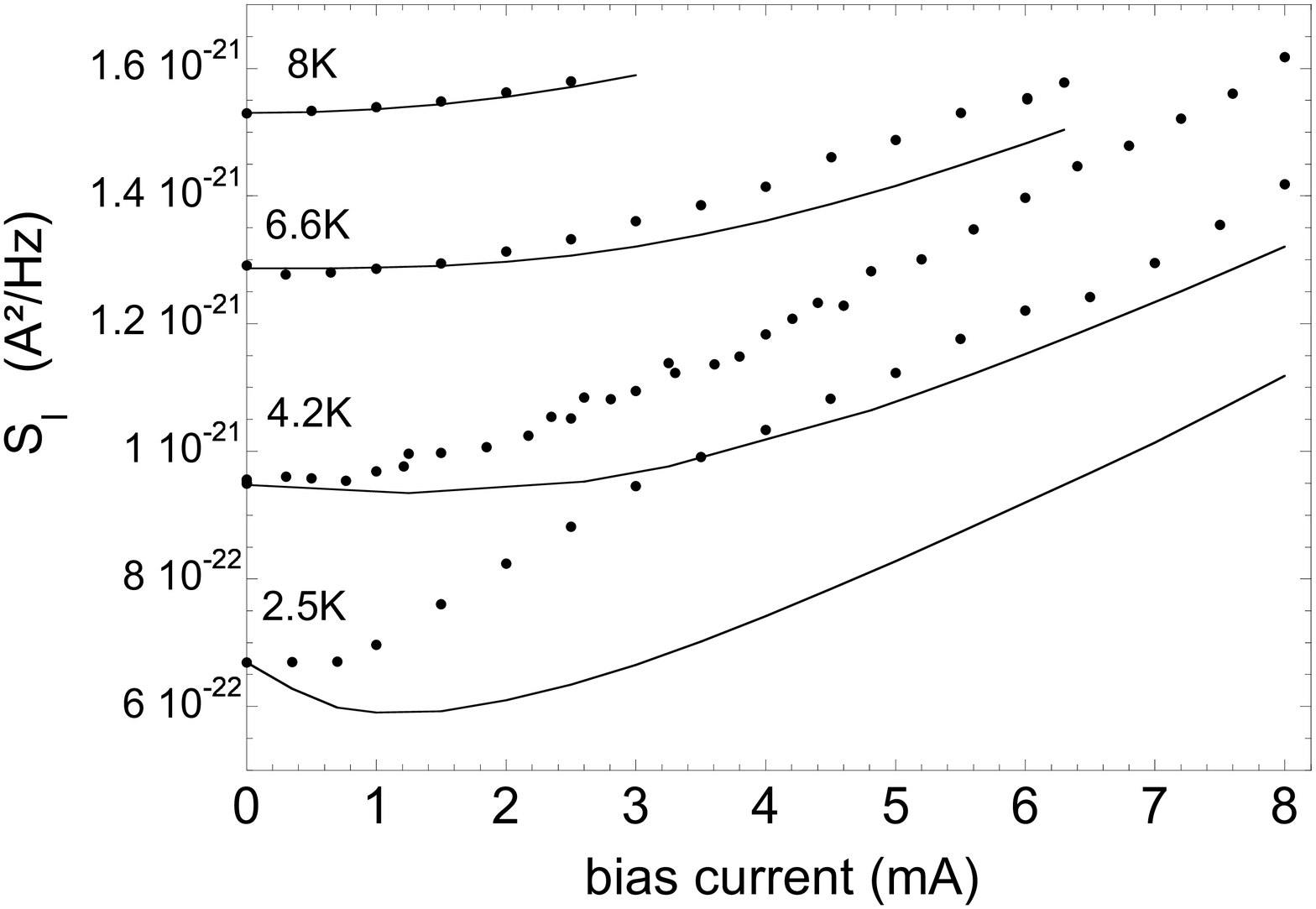,width=87mm}
\caption{Noise current density $S_{I}$ in the $L \approx 0.5 \mu m$ SNS sample at various temperatures (dots). The solid lines are the shot noise prediction for charges $e$ in a mesoscopic diffusive normal junction including the crossover to the thermal noise at zero bias. The noise follows this prediction only when superconductivity is weakened ($T \approx T_{c}=8.3K$), otherwise it is clearly higher.} 
\label{fig3jehl}
\end{figure}

 The shot noise results are presented in Fig. \ref{fig3jehl} for various temperatures. The Josephson coupling between the two superconducting banks is avoided by staying above $2K$: then the correlation length is substantially smaller than the sample length (typically $0.13\mu m$ at $2.5K$ and $0.5\mu m$). An exponentionally small Josephson coupling is important to study the low bias regime where the crossover between equilibrium (Johnson-Nyquist) and non-equilibrium (shot noise) regimes takes place ($e^{*}V \simeq 2k_{B}T$). The observation of this crossover has been a decisive argument in the study of fractional charges by noise measurements \cite{glattli}. A large Josephson coupling would also be responsible for another contribution to the current noise as shown in resistively shunted Josephson junctions \cite{koch}. The shot noise data at $8K$ (where superconductivity is already dramatically weakened at equilibrium) follow the solid line corresponding to the so called ${1\over 3}$ quantum shot noise suppression in normal mesoscopic diffusive samples \cite{buttiker}, including the thermal crossover: $S_{I}={2\over 3}[4k_{B}T/R_{d} + eIcoth(eV/2k_{B}T)]$ \cite{nagaev}. 

\begin{figure}
\psfig{figure=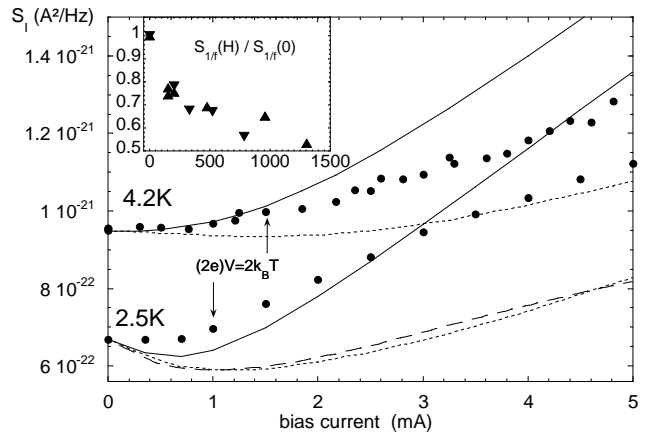,width=87mm}
\caption{Shot noise compared to predictions for the Normal case ($e^{*}=e$, dotted lines) and for the NS case ($e^{*}=2e$, solid lines) and to the heating estimation at $2.5K$ (dashed line, see text). The arrows indicate the thermal crossover for charges $2e$. Inset: amplitude of the flicker ($1/f$) noise as a function of magnetic field, normalized to the value at zero field and measured with DC bias current $3.2mA$ (lower triangles) and $3.9mA$ (upper triangles). The reduction by a factor of two occurs with a characteristic field related to $L_{\Phi}$.} 
\label{fig4jehl}
\end{figure}

Obviously the critical current at such temperatures is substantially smaller than at $4.2K$. Now as the temperature decreases the data coincide less and less with the normal prediction. As expected qualitatively the superconductivity is responsible for an increase in noise because it allows a new mechanism for charge transfer through the NS interfaces: the Andreev reflection of an electron as a hole and the transfer of a pair on the S side. We emphasize that, unlike in experimental (and theoretical) studies of short SNS systems in the coherent MAR regime, the Johnson value is always found at vanishingly small bias voltage and the crossover to the shot noise regime is smooth. The minimum observed at the onset of the $2.5K$ curve is a consequence of the peak in the differential conductance \cite{khlus}. In diffusive samples $L>l/\Gamma$ ($L$ is the length, $\Gamma$ the transparency of the NS interface and $l$ the elastic mean free path) the shot noise is expected to be doubled in NS samples compared to N samples \cite{beenakker}. In the asymptotic limit $eV \gg kT$ this means $S_{I}={2\over 3} (2e\langle I\rangle)$ instead of $S_{I}={1\over 3} (2e\langle I\rangle)$. In Fig. \ref{fig4jehl} we have plotted the equation given above for $S_{I}$ in the N case as well as the same equation with a charge $2e$ instead of $e$. We use this naive approach  because models describing the Johnson to shot noise crossover in NS are restricted to the one channel case only \cite{martin}. In this case however the exact calculation is close to the approximate $e\to 2e$ substitution.  We found good agreement between the data at $4.2K$ and $2.5K$ and the curves for a doubled charge $e^{*}=2e$, at low enough bias currents. We believe this is an experimental confirmation of the predicted doubled shot noise at NS interfaces.

\par Recent shot noise experiments gave rise to important discussions about heating effects. The crucial role of the reservoirs has been emphasized \cite{roukes}\cite{henny}. We calculated the thermal power that can be transferred through the NS interfaces by the single-particle excitations and the (thermal) noise associated with the hot electrons within the normal metal. In our temperature range ($T>2K$) the electron-phonon interaction is certainly able to restore the electrons closer to equilibrium. However as $L_{eph}\gtrsim L$ the contribution of the phonons is presumably too small to decrease the noise substantially\cite{naveh}, thus we neglected this mechanism in the heating calculation. We used the Andreev \cite{andreev} thermal resistance for the NS barrier to calculate the power that can be transferred through the NS interfaces. Then if we consider the power which is injected we obtain the temperature profile along the sample, taking into account both the Wiedemann-Franz law inside the normal part and the temperature jump across the NS interface due to its thermal resistance. Finally the noise due to these ``hot'' electrons is calculated with the Johnson noise formula. The result is plotted at $2.5K$ in Fig. \ref{fig4jehl} (dashed line). Clearly the Andreev thermal resistance gives an overheating effect higher than in the normal case. However at our relatively large temperatures this heating effect cannot quantitatively account for the data. At dilution refrigerator temperatures this heating effect becomes substantial as pointed out by Hoss et al \cite{strunk}. These authors used the Wexler formula and the BTK model \cite{btk} to account for electron heating. Our calculation uses the Andreev \cite{andreev} thermal resistance which contains no adjustable parameters, but using their arguments with reasonable assumptions for the NS resistance leads to similar results.

\par Another strong evidence for mesoscopic effects even at high bias currents is provided by the $1/f$ noise results \cite{feng}. First, we expressed the amplitude of the $1/f$ noise in terms of Hooge's law: $S_{I}/I^{2}=\alpha_{H}/Nf$ where $\alpha_{H}$ is the phenomenological Hooge parameter. Assuming a carrier density $N\approx 18 \times 10^{-22} cm^{-3}$ in aluminium, we found $\alpha_{H} \approx 10^{-3}$, in agreement with the range $10^{-5}$ to $10^{-1}$ given in the literature for thin films made with various materials. The model developed by Feng {\it et al.} shows that at low temperature the motion of a single scattering center is responsible for corrections to the conductance because of interference over $L_\phi$. A striking consequence is that under a weak magnetic field the amplitude of the $1/f$ noise is expected to be reduced by a factor of two \cite{feng}. This prediction has been verified in bismuth films  and semiconductors \cite{birge}. We performed this experiment and also found the universal reduction by a factor of two as shown in the inset of Fig. \ref{fig4jehl}. This result obtained for bias currents $3.2$ and $3.9 mA$ demonstrates that even at these high currents the mesoscopic features are conserved. Indeed the characteristic decay length over which the field is reduced is directly related to $L_{\phi}$. Stone \cite{feng} established that for a reduction by $75\%$, $H\approx 0.2L_{\Phi}^{2}h/e$. Using this relation we obtain for the two relevant bias currents $L_{\Phi}\approx 0.2\mu m$, i.e. a smaller value than inferred from weak localization measurements ($L_{\phi}\approx 0.8\mu m$). Nevertheless $L_{\Phi}$ remains comparable to $L$ even at high bias current. This result indicates that the inelastic lengths $L_{ee}$ or $L_{eph}$ (and therefore $L_{\phi}$) are not drastically reduced when several $mA$ are driven through the junction.

\par In summary, we performed the first $1/f$ and shot noise measurements in very low impedance SNS junctions in a high temperature regime which inhibits MAR features. We observed the shot noise enhancement due to Andreev reflections at NS interfaces. Under appropriate voltage and temperature conditions we see the predicted doubled shot noise due to the transfer of electron pairs through the NS boundaries. We estimated the thermal properties of the SNS structure with the Wiedemann-Franz law and the Andreev thermal conductance at the NS boundary and concluded that heating cannot be responsible for the observed noise. The reduction of the $1/f$ noise by a weak magnetic field demonstrates that the mesoscopic properties are not dramatically reduced by high currents. 
\\ We are grateful for fruitful discussions with C. Strunk, Y. Naveh, Th. Martin and V. Shumeiko.


\end{document}